\documentclass[sigconf]{acmart}
\usepackage{lmodern} 
\usepackage{lmodern} 
\AtBeginDocument{%
  }

\setcopyright{acmlicensed}
\copyrightyear{2026}
\acmYear{2026}
\acmDOI{XXXXXXX.XXXXXXX}
\acmConference[ICSE 2026]{2026 International Conference on Software Engineering}{April 12, 2026}{Rio de Janeiro, Brazil}
\acmISBN{978-1-4503-XXXX-X/2026/06}

\usepackage{subcaption}     
\definecolor{BrickRed}{rgb}{0.8, 0.25, 0.33}
\usepackage{tabularx}
\usepackage{enumitem}

\begin{document}

\title{Natural Language Summarization Enables Multi-Repository Bug Localization by LLMs in Microservice Architectures}

\author{Amirkia Rafiei Oskooei}
\orcid{0009-0004-3490-550X}
\email{amirkia.oskooei@intellica.net}
\email{amirkia.oskooei@std.yildiz.edu.tr}
\affiliation{%
  \institution{Intellica Business Intelligence Consultancy, R\&D Center}
  \institution{Yildiz Technical University, Dept. of Computer Eng.}
  \city{Istanbul}
  \country{Turkey}
}

\author{S. Selcan Yukcu}
\email{selcan.yukcu@intellica.net}
\affiliation{%
  \institution{Intellica Business Intelligence Consultancy, R\&D Center}
  \city{Istanbul}
  \country{Turkey}
}

\author{Mehmet Cevheri Bozoglan}
\email{cevheri.bozoglan@intellica.net}
\affiliation{%
  \institution{Intellica Business Intelligence Consultancy, R\&D Center}
  \city{Istanbul}
  \country{Turkey}
}

\author{Mehmet S. Aktas}
\orcid{0000-0001-7908-5067}
\email{aktas@yildiz.edu.tr}
\affiliation{%
  \institution{Yildiz Technical University, Department of Computer Engineering}
  \city{Istanbul}
  \country{Turkey}
}

\renewcommand{\shortauthors}{Rafiei Oskooei et al.}

\begin{abstract}
Bug localization in multi-repository microservice architectures is challenging due to the semantic gap between natural language bug reports and code, LLM context limitations, and the need to first identify the correct repository. We propose reframing this as a natural language reasoning task by transforming codebases into hierarchical NL summaries and performing NL-to-NL search instead of cross-modal retrieval.
Our approach builds context-aware summaries at file, directory, and repository levels, then uses a two-phase search: first routing bug reports to relevant repositories, then performing top-down localization within those repositories. Evaluated on DNext, an industrial system with 46 repositories and 1.1M lines of code, our method achieves Pass@10 of 0.82 and MRR of 0.50, significantly outperforming retrieval baselines and agentic RAG systems like GitHub Copilot and Cursor.
This work demonstrates that engineered natural language representations can be more effective than raw source code for scalable bug localization, providing an interpretable repository $\rightarrow$ directory $\rightarrow$ file search path, which is vital for building trust in enterprise AI tools by providing essential transparency.
\end{abstract}



\begin{teaserfigure}
    \centering
    \includegraphics[width=\textwidth]{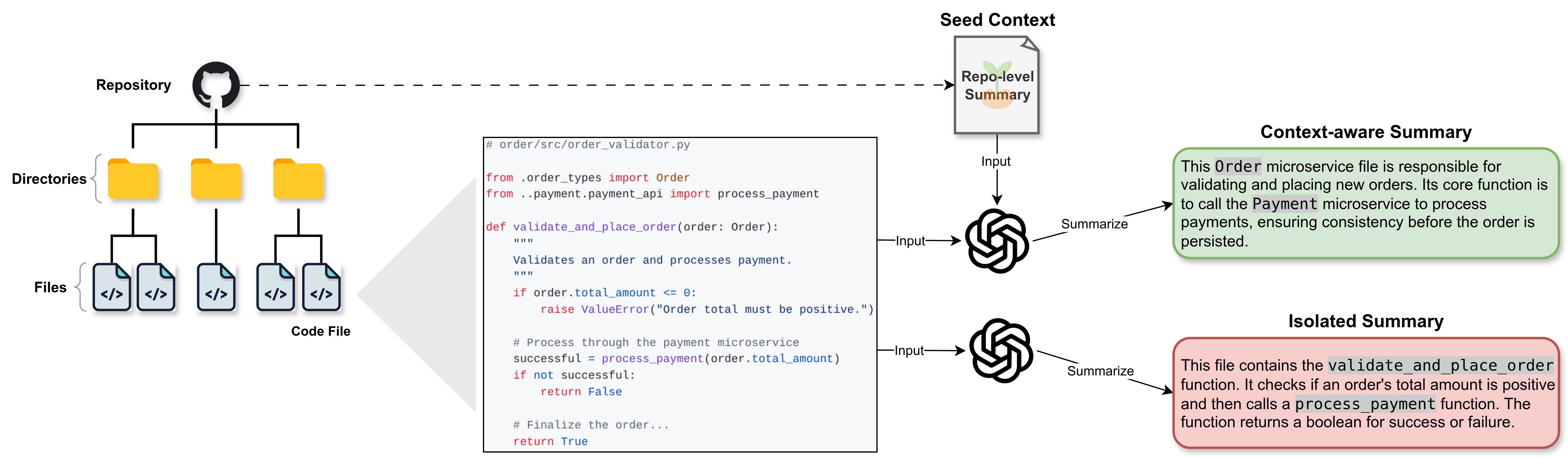}
    \caption{A comparison of isolated vs. context-aware summarization. \textbf{(Red)} An isolated summary describes a file's low-level implementation details. \textbf{(Green)} Our context-aware approach, primed with a repository-level "seed context," produces a summary that explains the file's architectural role and purpose, which is more effective for bug localization.}
    \Description{A diagram comparing isolated and context-aware summarization, highlighting differences in summary detail and effectiveness for bug localization.}
    \label{fig:context-aware}
\end{teaserfigure}


\maketitle

\section{Introduction}
\label{sec:introduction}

Large Language Models (LLMs) are increasingly applied to complex software engineering tasks like bug localization. However, current methods, which primarily rely on Retrieval-Augmented Generation (RAG) over raw source code, face significant challenges in large-scale, industrial settings. These systems often struggle with three key limitations.
First, they must bridge the \textit{semantic gap}, where the natural language of a user's bug report often lacks semantic overlap with the technical terms in the code. 
Second, they are constrained by the LLM's \textit{finite context window}, which prevents them from comprehending an entire codebase holistically. 
Third, they are often designed with an implicit \textit{single-repository assumption}. This makes them ineffective for large-scale software projects composed of many interacting microservices, as they lack a mechanism to first route a bug report to the correct repository (codebase) before localization can begin.

\begin{figure}[t]
    \centering
    \includegraphics[width=0.9\columnwidth]{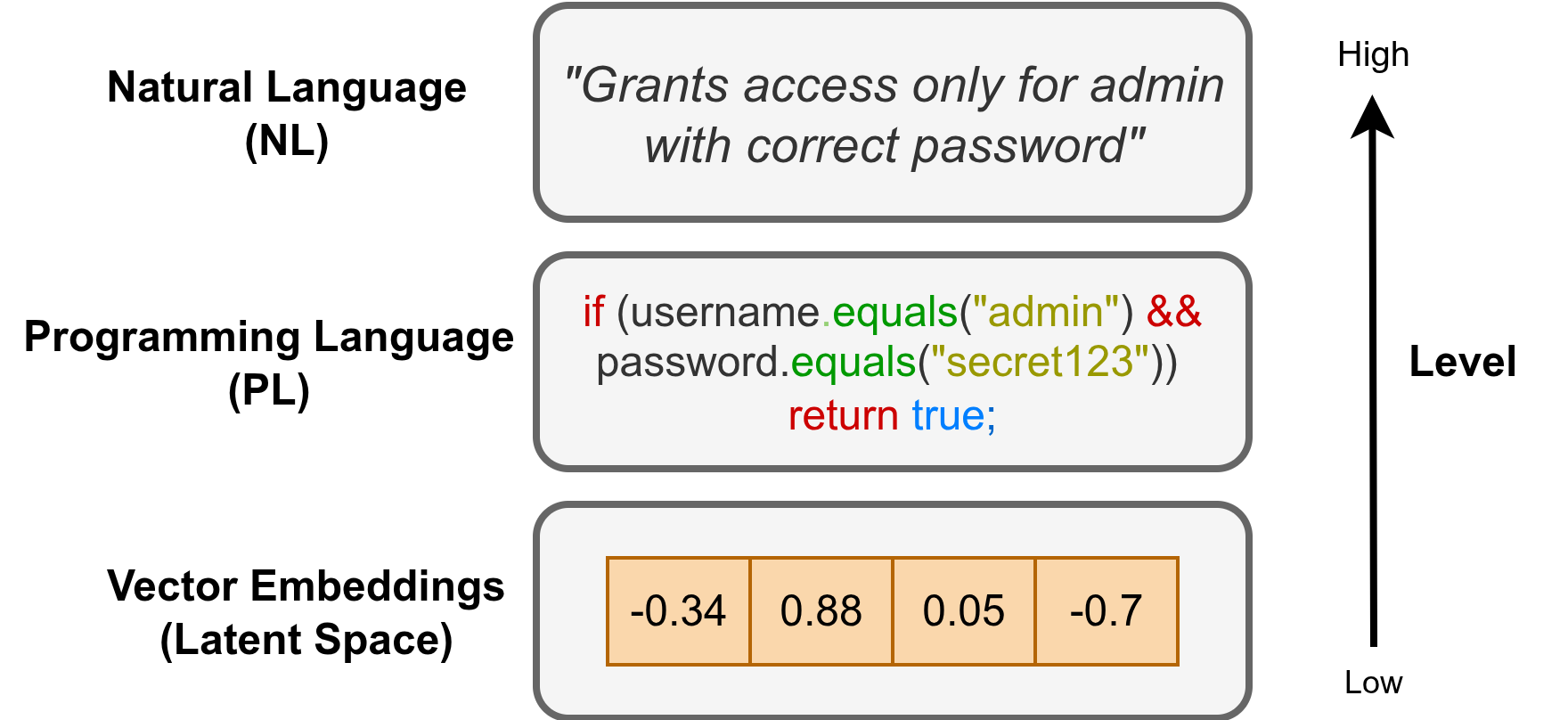}
    \caption{The levels of code abstraction. While traditional methods move down from Programming Language (PL) to low-level vector embeddings, our approach moves up, transforming PL into a higher-level Natural Language (NL) representation to enable semantic reasoning with LLMs.}
    \Description{A diagram showing three levels of code representation. At the bottom are vector embeddings, in the middle is programming language code, and at the top is a natural language summary. An arrow indicates that traditional methods move down this ladder, while our approach moves up.}
    \label{fig:abstraction_levels}
\end{figure}

In this work, we explore an alternative paradigm. As illustrated in Figure~\ref{fig:abstraction_levels}, most current methods move \textit{down} the abstraction ladder, transforming source code (PL) into low-level vector embeddings for similarity search. We hypothesize that for complex bug localization, it is more effective to move \textit{up} the ladder. By representing code at a higher level of abstraction—natural language (NL) summaries—we reframe the problem. This shift from a cross-modal retrieval task to a unified NL-to-NL reasoning task is designed to directly leverage the powerful semantic understanding of modern LLMs.

We present a methodology to realize this approach. To manage the scale of enterprise systems, we first transform the source code across all repositories within a project into a compact, hierarchical NL knowledge base. This is achieved through \textit{context-aware summarization}, which creates semantically rich descriptions that capture the architectural purpose of each component. This compact representation helps mitigate the LLM's context window limitations. The knowledge base is then navigated using a scalable, \textit{two-phase hierarchical search} that first identifies the most probable repository before performing a targeted, top-down search for the faulty file.

Our primary contributions are:
\begin{itemize}
    \item A methodology for transforming a multi-repository software project into a hierarchical NL knowledge base, enabling a scalable, end-to-end bug localization workflow.
    \item A two-phase search framework that explicitly addresses the multi-repository routing problem, a critical challenge for bug localization in microservice architectures.
    \item A comprehensive empirical evaluation on a large-scale industrial project, demonstrating the effectiveness of our approach over strong RAG and retrieval-based baselines.
    \item An interpretable and auditable reasoning process (repository $\rightarrow$ directory $\rightarrow$ file), which can enhance developer trust in AI-powered tools.
\end{itemize}
\section{Methodology}
\label{sec:methodology}

To address the challenges of scale, context, and architectural complexity outlined in Section~\ref{sec:introduction}, our methodology is designed as a two-stage process. First, we transform the entire multi-repository codebase into a compact and semantically rich \textbf{NL Knowledge Base}. Second, we employ a \textbf{Two-Phase Search} to efficiently navigate this knowledge base and pinpoint the source of a bug. This approach systematically converts a cross-modal search problem into a more effective NL-to-NL reasoning task.

\subsection{Building Knowledge Base}
\label{sec:build_kb}

The primary obstacle to applying LLMs on enterprise-scale systems is the finite context window. To overcome this, we construct a hierarchical representation of the codebase that is both compact and information-rich. This offline process involves two key steps.

\begin{figure*}[h]
    \centering
    \includegraphics[width=0.75\textwidth]{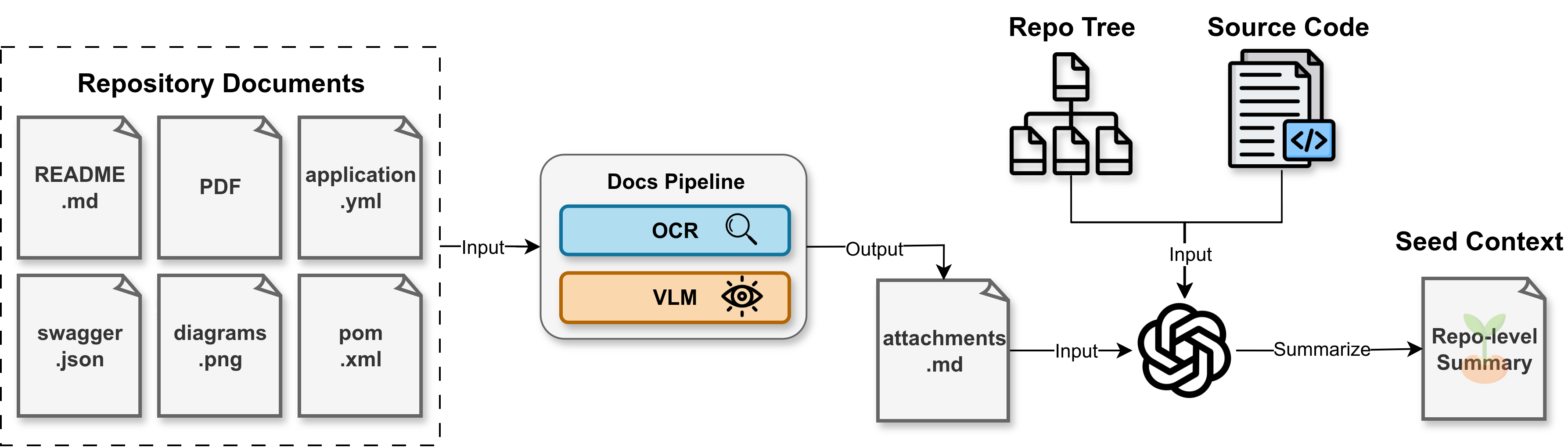}
    \caption{The generation process for the repository-level summary. A multimodal pipeline processes various repository documents into a single text format. These attachments, along with the repository tree and source code, are used to prompt an LLM to create a comprehensive summary that serves as the seed context for all downstream tasks.}
    \Description{A diagram showing the process of see context generation}
    \label{fig:seed-plus-doc}
\end{figure*}

\paragraph{Repository-Level Seed Context.}
The process begins by generating a high-quality, repository-level summary for each microservice, which serves as a "seed context." This holistic summary ensures that lower-level descriptions understand their architectural role, a critical distinction shown in Figure~\ref{fig:context-aware}. To generate this seed context, we prompt an LLM with a comprehensive set of artifacts, a process detailed in Figure~\ref{fig:seed-plus-doc}:
\begin{itemize}
    \item \textbf{The Repository Tree:} The directory and file structure.
    \item \textbf{Source Code:} A linearized version of the codebase (code snippets).
    \item \textbf{Processed Attachments:} Project documentation, including READMEs and descriptive text extracted from diagrams (e.g., UML, ER) using a multi-modal OCR and Vision-Language Model (VLM) pipeline.
\end{itemize}

\begin{figure}[!h]
    \centering
    \includegraphics[width=0.75\columnwidth]{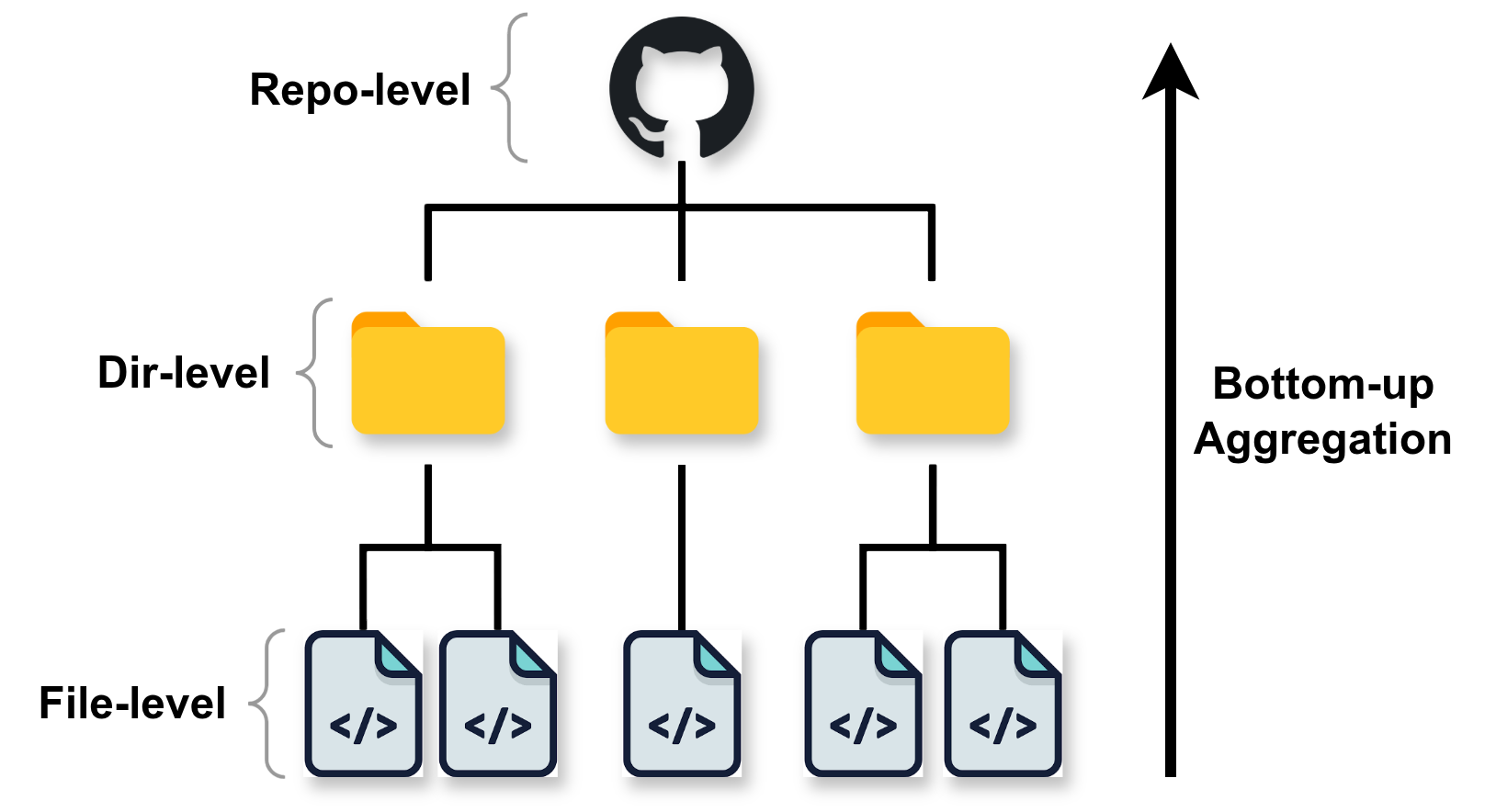}
    \caption{Bottom-up construction of the knowledge tree. Leaf nodes (file-level summaries) are generated first, using the repository-level summary as seed context. These are then aggregated to create intermediate nodes (directory-level summaries), forming a hierarchical NL representation of the codebase.}
    \Description{A diagram showing the code repository structure}
    \label{fig:bottom-up}
\end{figure}

\paragraph{Bottom-Up Aggregation.}
With the seed context established, we then construct the knowledge tree in a bottom-up fashion (Figure~\ref{fig:bottom-up}). For each source file, an LLM is prompted with its raw code \textit{and} the repository's seed context to generate a concise, context-aware summary. These file-level summaries (the leaves of our tree) are then aggregated at the directory level. For each major directory (e.g., \texttt{/service}, \texttt{/controller}), its constituent file summaries are provided to the LLM to generate a coherent directory-level summary, forming the intermediate nodes of the knowledge tree.

\subsection{Two-Phase Search}
\label{sec:search_kb}

During inference, we leverage the NL knowledge base to perform an efficient top-down search. This process is explicitly designed to handle the multi-repository nature of microservice architectures.

\begin{figure}[!h]
    \centering
    \includegraphics[width=0.6\columnwidth]{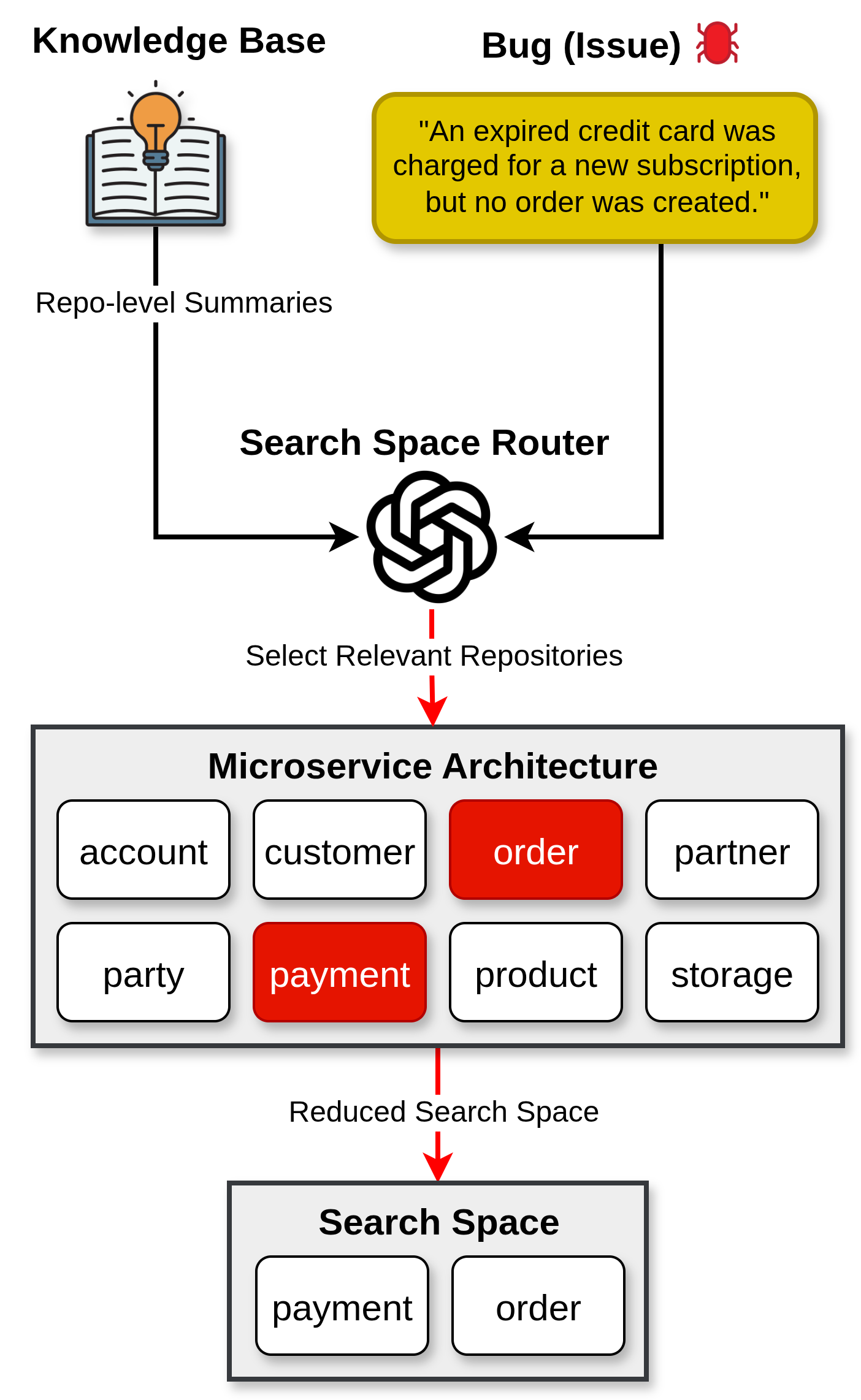}
    \caption{The Search Space Routing process. The LLM compares a bug report against repository-level summaries to identify a ranked list of candidate microservices, focusing the search.}
    \Description{A diagram demonstrating the process of search space routing}
    \label{fig:space-router-small}
\end{figure}

\paragraph{Phase 1: Search Space Routing.}
Given a bug report, the first challenge is to identify the correct microservice (repository). As illustrated in Figure~\ref{fig:space-router-small}, our framework routes the query by prompting an LLM to compare the bug report against the high-level, repository-level summaries of all microservices. This produces a ranked list of candidate repositories, ensuring the subsequent search is focused only on relevant code repositories.

\paragraph{Phase 2: Top-Down Bug Localization.}
Within each candidate repository, we perform a two-step hierarchical search to find the specific buggy file (Figure~\ref{fig:bug-localization}). First, the LLM performs \textit{Directory-Level Filtering} by comparing the bug report against all directory-level summaries to identify the top-$k$ most relevant directories. Second, the search space is narrowed to the files within these directories for a final \textit{File-Level Ranking}. The LLM performs a detailed analysis of these filtered file summaries to produce the final ranked list of files most likely to be the source of the bug.

\begin{figure}[!h]
    \centering
    \includegraphics[width=0.75\columnwidth]{figures/bug-localization.png}
    \caption{The top-down bug localization process. The search is first narrowed to candidate directories \textbf{(Upper Red Rectangle)} before a focused search over file-level summaries finds the exact bug location \textbf{(Lower Red Rectangle)}.}
    \Description{A diagram showing the process of bug localization}
    \label{fig:bug-localization}
\end{figure}

This structured \texttt{repository} $\rightarrow$ \texttt{directory} $\rightarrow$ \texttt{file} search path provides an inherently transparent and auditable reasoning process.\footnote{To enhance the quality of reasoning at each stage, all prompts used during the inference process explicitly leverage Chain-of-Thought (CoT) prompting techniques.} By decomposing the localization task into a series of explicit, verifiable steps, our method addresses the "black box" nature of many end-to-end systems, contributing to the goal of \textbf{Explainable AI} in developer tools.
\section{Experiments and Evaluation}
\label{sec:experiments}

We conduct an empirical study to evaluate our proposed architecture for LLM-based bug localization. Our evaluation aims to answer two primary questions: (1) How effectively does our NL-to-NL approach perform against state-of-the-art baselines on a complex, multi-repository benchmark? (2) How critical is the hierarchical search component to the method's accuracy and efficiency?

\subsection{Experimental Setup}

\paragraph{Dataset.}
Our evaluation is performed on \textbf{DNext}~\citep{dnext-technology}, a proprietary, industrial-scale microservice system for the telecommunications sector. Its scale (see Table~\ref{tab:dataset-stats}) and use of real-world, often noisy, bug reports provide a challenging benchmark that standard academic datasets cannot replicate. The ground truth for each bug ticket is the set of files modified in the pull request that resolved the issue.

\begin{table}[h]
\centering
\caption{Statistics of the DNext microservice dataset.}
\label{tab:dataset-stats}
\begin{tabular}{@{}ll@{}}
\toprule
\textbf{Metric} & \textbf{Value} \\ \midrule
Programming Language & Java \\
Number of Repositories & 46 \\
Total Number of Code Files & 7,077 \\
Total Physical Lines of Code & $\sim$1.1M \\ \midrule
Number of Bug Tickets & 87 \\
Average Buggy Files per Ticket & 7.2 \\ \bottomrule
\end{tabular}
\end{table}

\paragraph{Baselines.}
We compare our method against two strong categories of baselines. For all LLM-based methods, including our own, we use the \textbf{gpt-4.1} model to ensure a fair comparison of reasoning capabilities.
\begin{itemize}
    \item \textbf{Flat Retrieval:} We use traditional IR baselines, \textbf{UniXcoder}~\citep{guo2022unixcoder} and \textbf{GraphCodeBERT}~\citep{guo2020graphcodebert}, to retrieve the most similar code files based on cosine similarity between embeddings.
    \item \textbf{Agentic RAG:} We evaluate against state-of-the-art, repository-aware code assistants that operate on source code via RAG. We use two leading commercial systems: \textbf{GitHub Copilot}~\citep{github-copilot} and \textbf{Cursor}~\citep{cursor-ai}.
\end{itemize}

\paragraph{Metrics.}
We evaluate performance using standard metrics: \textbf{Pass@k} and \textbf{Recall@k}. \textbf{Pass@k} measures the proportion of bug reports where \textit{at least one} correct file is found in the top-$k$ results. \textbf{Recall@k} measures the fraction of \textit{all} correct files for a given bug that are successfully retrieved within the top-$k$ results. Given that the maximum number of modified files for any bug in our dataset is 10 (average 7.2), we set $k=10$ as a fair and comprehensive threshold that covers the full scope of a typical fix without artificially inflating scores with a large retrieval window. For the preliminary search space routing phase, we use a tighter $k=3$.

\subsection{Main Results}

\paragraph{Search Space Routing Performance.}
Table~\ref{tab:routing-results} shows our method's effectiveness at the critical first step: routing a bug to the correct microservice. Our summary-based router outperforms the Agentic RAG baselines on coverage metrics (Pass@3 and Recall@3). This is a crucial advantage for ambiguous bugs that may span multiple services, as our method is more likely to include all relevant repositories in the initial search space.

\begin{table}[h]
\centering
\caption{Search Space Routing performance. Our NL-to-NL approach is more effective at identifying the correct set of candidate repositories.}
\label{tab:routing-results}
\begin{tabular}{@{}lccc@{}}
\toprule
\textbf{Approach} & \textbf{Pass@3} & \textbf{Recall@3} & \textbf{MRR}  \\ \midrule
\textit{Flat Retrieval} & N/A & N/A & N/A \\ 
\textit{Agentic RAG} & & \\
\quad GitHub Copilot &  0.82 & 0.82 & \textbf{0.77} \\
\quad Cursor &  0.91 & 0.91 & 0.72 \\ \midrule
\textbf{Ours} & \textbf{0.93} & \textbf{0.93} & 0.67 \\ \bottomrule
\end{tabular}
\end{table}

\paragraph{Bug Localization Performance.}
Table~\ref{tab:search-results} shows the end-to-end file localization performance. Our hierarchical approach achieves the highest Pass@10 and MRR, indicating it is both more likely to find a correct file and to rank the first correct file higher than any baseline. It also matches the best-performing baseline (Cursor) on Recall@10, the metric most sensitive to multi-file bugs. These results validate our central hypothesis: a structured search over an NL knowledge base can be a more accurate and effective paradigm than retrieving from raw source code in some use cases.

\begin{table}[h]
\centering
\caption{File-level bug localization performance. Our approach outperforms both retrieval and Agentic RAG baselines.}
\label{tab:search-results}
\small
\begin{tabular}{@{}lccc@{}}
\toprule
\textbf{Approach} & \textbf{Pass@10} & \textbf{Recall@10} & \textbf{MRR} \\ \midrule
\textit{Flat Retrieval} & & & \\
\quad UniXcoder & 0.23 & 0.16 & 0.14\\ 
\quad GraphCodeBERT & 0.62 & 0.34 & 0.18 \\ \midrule
\textit{Agentic RAG} & & & \\
\quad GitHub Copilot & 0.61 & 0.27 & 0.25 \\
\quad Cursor & 0.57 & \textbf{0.49} & 0.27 \\ \midrule
\textbf{Ours (Hierarchical)} & \textbf{0.82} & \textbf{0.49} & \textbf{0.50} \\ \bottomrule
\end{tabular}
\end{table}

\subsection{Ablation Study: The Impact of Hierarchy}
\label{sec:ablation}

To validate our claim that the hierarchical search is a critical component for both accuracy and efficiency, we conduct an ablation study. We compare our full method against an ablated version, which omits the directory-level filtering step. In this version, after routing to the correct repository, the LLM is prompted with the bug report and the summaries of \textit{all} files within that repository to produce a final ranking directly.

\begin{table}[h]
\centering
\caption{Ablation study on the impact of hierarchical search. The `w/o filtering` row shows the significant performance degradation and cost increase when the directory-level filtering is removed.}
\label{tab:ablation-results}
\begin{tabularx}{\columnwidth}{@{}l *{4}{>{\centering\arraybackslash}X} @{}}
\toprule
\textbf{Method} & \textbf{Pass@10} & \textbf{Recall@10} & \textbf{MRR} & \textbf{Avg. Tokens} \\ \midrule
\textbf{w/ filtering} & \textbf{0.82} & \textbf{0.49} & \textbf{0.50} & \textbf{~37K + ~38K} \\
\textbf{w/o filtering} & 0.67 \small\textcolor{BrickRed}{(-18\%)} & 0.31 \small\textcolor{BrickRed}{(-37\%)} & 0.23 \small\textcolor{BrickRed}{(-54\%)} & ~108K \small\textcolor{BrickRed}{(+44\%)} \\ \bottomrule
\end{tabularx}
\end{table}

The results of our ablation study, presented in Table~\ref{tab:ablation-results}, powerfully validate our central architectural claim. Removing the directory-level filtering step forces the LLM to solve a classic "needle-in-a-haystack" problem, searching for a few relevant files within the noisy context of an entire repository's summaries. This leads to a catastrophic drop in performance across all metrics, with MRR plummeting by 54\% and Recall@10 falling by 37\%. In contrast, our hierarchical method acts as a cognitive scaffold, breaking the complex localization task into manageable steps to focus the LLM's reasoning. This approach is not only substantially more accurate but also more scalable, as the ablated version requires a 44\% increase in prompt tokens, making it prohibitively expensive and proving that our hierarchical design is critical for both performance and cost-effectiveness.

\subsection{Qualitative Analysis}
To understand \textit{why} our NL-to-NL approach succeeds where retrieval fails, we present two case studies in Figure \ref{fig:qualitative_analysis}. These examples show how reasoning over the \textbf{conceptual purpose} of code, rather than just its keywords, allows our method to solve complex bugs.

\begin{figure}[h!]
\begin{minipage}{\linewidth}
\hrule
\vspace{2mm}
\noindent\textbf{Case Study 1: Conditional Logic Bug}
\vspace{2mm} 

\small
\noindent \textbf{Bug Report:} \small\textit{"Revision number is not increased in certain cases. External Reference is added but revision remains as zero. However, a PATCH request to replace the `/name` field does increase the revision... This should be fixed for financialAccount and partyAccount}

\vspace{2mm}

\small
\noindent \textbf{The Challenge:} A successful tool must understand the code's conditional business logic—why one PATCH operation behaves differently from another—not just find files related to "Account".

\vspace{1mm}

\noindent\textbf{Baseline Failure:}
The Agentic RAG retrieved generic files, such as:
\begin{itemize}[leftmargin=*]
    \item \texttt{controller/PartyAccountController.java}
    \item \texttt{model/FinancialAccount.java}
\end{itemize}

\noindent \textit{Analysis: This approach completely missed the nuanced validation logic central to the bug.}

\vspace{2mm}

\noindent\textbf{Our Method's Success:}
Our method identified the crucial service classes and, most importantly, the specific validator governing the patch operation:
\begin{itemize}[leftmargin=*]
    \item \texttt{service/FinancialAccountService.java}
    \item \texttt{validation/PartyAccountJsonPatch...Validator.java}
\end{itemize}
\noindent \textit{Analysis: By understanding the file's architectural role, it pinpointed the exact component responsible for the faulty logic.}
\vspace{2mm}
\hrule
\vspace{2mm}
\hrule
\vspace{2mm}


\noindent\textbf{Case Study 2: Referential Integrity Bug}
\vspace{2mm}

\small
\noindent \textbf{Bug Report:} \small\textit{"There's a data integrity issue in order management. The system is allowing us to create new service orders that refer to service specifications that don't actually exist in the catalog..."}

\vspace{2mm} 

\small
\noindent \textbf{The Challenge:} The bug is caused by a missing validation check between two different business domains. The tool must reason about the expected interaction, not just search for existing code.

\vspace{2mm}

\noindent \textbf{Baseline Failure:}
The Agentic RAG identified the general topic (`ServiceOrder`) but returned irrelevant mappers and high-level application files:
\begin{itemize}[leftmargin=*]
    \item \texttt{controller/ServiceOrderController.java}
    \item \texttt{mapper/ServiceOrderMapper.java}
\end{itemize}
\noindent  \textit{Analysis: It found the right topic but could not diagnose the specific, missing integrity check.}

\vspace{2mm}

\noindent\textbf{Our Method's Success:}
Our method correctly identified the specialized validator classes where the missing logic should have been implemented:
\begin{itemize}[leftmargin=*]
    \item \texttt{validators/OrderItemServiceSpec...PatchValidator.java}
    \item \texttt{util/ServiceSpecificationExistenceValidatorHelper.java}
\end{itemize}
\noindent \textit{Analysis: It succeeded by reasoning about the required interaction between system components, even when the code for it was absent.}
\vspace{2mm}
\hrule
\end{minipage} 
\caption{Qualitative analysis of two challenging bugs. Our method succeeds by reasoning about conceptual logic, while the baseline fails by relying on superficial keyword matching.}
\label{fig:qualitative_analysis}
\end{figure}
\section{Related Works}
\label{sec:related_works}

Our work is situated at the intersection of code retrieval, LLM-based bug localization, and code summarization.

\paragraph{NL-to-PL Code Retrieval.}
The foundational challenge in bug localization is bridging the cross-modal gap between natural language (NL) and programming language (PL). Foundational models like CodeBERT~\citep{feng2020codebert}, UniXcoder~\citep{guo2022unixcoder}, and GraphCodeBERT~\citep{guo2020graphcodebert} established the viability of joint NL-PL embeddings by incorporating code structure. While subsequent work has sought to mitigate the persistent vocabulary mismatch through techniques like dynamic execution features~\citep{martinez2024improving} or NL-augmented retrieval~\citep{chen2024code}, these methods still operate on the difficult cross-modal boundary our NL-to-NL approach is designed to circumvent.

\paragraph{LLM-based Bug Localization.}
Modern systems leverage LLMs for more sophisticated reasoning. Strategies include augmenting the query or code before retrieval~\citep{jain2024llm, li2024rewriting}, combining retrieval with iterative codebase analysis~\citep{asad2025leveraging}, or leveraging structured context via external memory~\citep{yeo2025improving} and code graphs~\citep{liu2024graphcoder}. State-of-the-art agentic frameworks like OrcaLoca~\citep{yu2025orcaloca} and COSIL~\citep{jiang2025cosil} demonstrate impressive navigation, and hierarchical approaches like BugCerberus~\citep{chang2025bridging} use specialized models at different granularities. However, BugCerberus focuses on vertical granularity (File $\to$ Statement) within a single repository context. Despite their power, these frameworks are architecturally constrained to \textit{single-repository settings}. They fundamentally lack a mechanism for multi-repository routing, making them unsuitable for modern microservice architectures where identifying the correct repository is the critical first step. Our two-phase framework directly addresses this critical gap.

\paragraph{Automatic Code Summarization.}
The field of automatic code summarization~\citep{sun2024source} provides the foundation for our representation engineering. Research has shown that code search is significantly improved when code snippets are paired with NL descriptions~\citep{heyman2020neural}. Recent advances have moved beyond method-level summaries to tackle higher-level units using hierarchical~\citep{sounthiraraj2025code, dhulshette2025hierarchical, sun2025commenting, oskooei2025repository} and context-aware~\citep{su2025context} generation strategies. While our previous work~\citep{oskooei2025repository} demonstrated the feasibility of hierarchical summarization for single-repository contexts, this paper extends that foundation into a comprehensive framework for multi-repository microservice architectures. We introduce the \textit{Search Space Router} to solve the critical "missing link" of repository identification and enhance the summarization strategy with context-aware propagation. Critically, retrieval over NL summaries has been shown to be a viable alternative to searching over code~\citep{tawosia2025meta}. While this body of work typically treats summaries as a documentation end-product, our key contribution is to systematically engineer these summaries into a task-specific \textbf{intermediate representation}. By doing so, we transform bug localization from a difficult cross-modal retrieval problem into a more tractable NL-to-NL reasoning task optimized for scale and architectural complexity.

\paragraph{Positioning Our Work} Our contribution is therefore not a new summarization model, but rather a novel act of \textbf{representation engineering}. We systematically apply established summarization techniques to build a purpose-built NL knowledge base, an intermediate representation designed specifically for the task of bug localization (which is a re-usable asset for other use-cases as mentioned in Section \ref{sec:discussion}). This architectural pattern is the key that unlocks two distinct advantages over existing methods: it provides the holistic context necessary to solve the critical \textbf{multi-repository} routing problem, and it reframes the task into a unified \textbf{NL-to-NL} reasoning challenge that better leverages the semantic power of LLMs. Consequently, our framework is not only more scalable for complex \textbf{microservice} environments but also inherently more \textbf{interpretable}, offering a non-black-box reasoning path that is critical for enterprise adoption where developer trust is paramount.
\section{Discussion}
\label{sec:discussion}

Our empirical results validate our central hypothesis: for bug localization in complex systems, an engineered natural language representation of source code can be sometimes more effective than the code itself. Our findings, including the ablation study, confirm that the hierarchical search over this representation is not only more accurate but also more cost-effective. This section discusses the broader implications of this architectural shift and acknowledges the limitations of our work.

\paragraph{Implications for AI-Powered Developer Tools.}
This work has several practical implications for the design of future SE tools:
\begin{itemize}
    \item \textbf{A Scalable Architecture for Microservices:} The single-repository assumption of most current tools is a critical architectural limitation. Our two-phase framework offers a concrete, scalable solution for applying LLM-based reasoning to enterprise-scale, multi-repository systems—a ubiquitous industrial paradigm.
    
    \item \textbf{A New Paradigm via Reusable Knowledge Bases:} The hierarchical NL knowledge base is a reusable asset. Beyond bug localization, it can power a new class of developer tools for tasks like conceptual code search, automated onboarding of new engineers, and architectural analysis, similar to tools such as DeepWiki \cite{deepwiki}. Our system is also capable of integrating with current code agents, allowing them to collaborate in identifying the location of the bugs and then resolving them.
    
    \item \textbf{Enhancing Trust through Interpretability:} The ``black box'' nature of many AI tools is a barrier to enterprise adoption. Our method's transparent reasoning path (\texttt{repository} $\rightarrow$ \texttt{directory} $\rightarrow$ \texttt{file}) is auditable by developers, fostering the trust required for effective human-AI collaboration in high-stakes environments, as discussed by \citep{korbak2025chain}.
\end{itemize}

\paragraph{Limitations and Future Work.}
Our study has two primary limitations. First, the initial generation of the knowledge base is computationally intensive.
As detailed in Appendix~\ref{app:cost}, while the initial construction of the hierarchical NL knowledge base is a one-time offline process, it introduces a non-negligible computational cost ($\approx$\$80 for DNext).  
In our industrial integration, this process is performed after each major release rather than after every commit. 
This scheduling choice is motivated by the nature of the generated summaries: since they capture the \textit{high-level purpose and architectural role} of files rather than their exact implementation details, rapid implementation changes rarely invalidate them.
However, as projects scale, both the \textbf{cost} and the \textbf{maintenance scheduling algorithm} for updating these summaries should be addressed in future work to improve efficiency and sustainability.

Second, our evaluation, while substantial, was conducted on a single large-scale industrial Java project. Validating the approach on projects in other languages (e.g., Python, Go) and architectural patterns (e.g., monolithic architectures) is an important next step for assessing generalizability.
\section{Conclusion}
\label{sec:conclusion}

Bug localization in modern, multi-repository software systems is a critical bottleneck where existing RAG-based tools fail due to architectural and context limitations. In this paper, we introduced a novel framework that reframes this challenge as a tractable NL-to-NL reasoning task. By systematically engineering source code into a hierarchical NL knowledge base and navigating it with a scalable, two-phase search, our approach improves upon strong RAG and retrieval baselines on a large-scale industrial benchmark. Ultimately, this work demonstrates the power of representation engineering in software engineering. By shifting the paradigm from retrieving over raw code to reasoning over a curated, human-understandable knowledge base, we open a promising new direction for AI-powered developer tools that are more scalable, accurate, and trustworthy.

\begin{acks}
This work was conducted at the R\&D Center of Intellica Business Intelligence Consultancy and was supported by the Scientific and Technological Research Council of Turkey (TÜBİTAK) under the TEYDEB 1501 program (Grant No. 3240105). 

A generative AI tool (Gemini 2.5 Pro) was used solely to assist with grammar and clarity improvements; the authors bear full responsibility for the scientific content and integrity of this work.
\end{acks}

\bibliographystyle{ACM-Reference-Format}
\bibliography{references}

\appendix
\section{Commercial Tool Prompting Strategy}
\label{sec:appendix_baselines}

To ensure a fair and reproducible comparison against the Agentic RAG baselines (GitHub Copilot and Cursor), we followed a standardized prompting procedure. Figures \ref{fig:baseline-file-prompt} and \ref{fig:baseline-repo-prompt} detail the exact setup and prompt templates used for both the multi-repository routing and file-level localization tasks.

\begin{figure}[tb]
    \centering
    \fbox{ 
        \begin{minipage}{0.93\columnwidth} 
            \textbf{Setup:} The target microservice (repository) is opened as the root folder in the IDE (VS Code with GitHub Copilot / Cursor). The agent's context is therefore limited to that single repository.
            \vspace{2mm}
            \hrule
            \vspace{2mm}
            \textbf{Prompt:} You are a bug localizer. Given the bug description below, rank the top-10 files that likely cause the bug. Rank them in order of relevance.
        \end{minipage}
    } 
    \caption{Prompt template for file-level bug localization using Agentic RAG baselines.}
    \label{fig:baseline-file-prompt}
\end{figure}

\begin{figure}[tb]
    \centering
    \fbox{ 
        \begin{minipage}{0.93\columnwidth} 
            \textbf{Setup:} All 46 microservice repositories are opened together in a single workspace from their shared root directory. This allows the agent to see all folders simultaneously.
            \vspace{2mm}
            \hrule
            \vspace{2mm}
            \textbf{Prompt:} You are a multi-repository bug localizer, selecting a search space within a large microservice architecture. Each microservice is represented as a folder in workspace. Given the bug description below, rank the top-3 microservices (folders) that likely cause the bug. Rank them in order of relevance.
        \end{minipage}
    } 
    \caption{Prompt template for multi-repository routing (search space selection) using Agentic RAG baselines.}
    \label{fig:baseline-repo-prompt}
\end{figure}


\section{Version Information}
\label{sec:appendix_versions}

All interactions with the Agentic RAG baselines were conducted using their respective interactive ``Chat'' or ``Ask'' modes to ensure the full reasoning capabilities of the underlying models were engaged. The specific versions used in our experiments were:
\begin{itemize}
    \item \textbf{GitHub Copilot:} We utilized version v1.104 (release date: August 29, 2025) of the VS Code extension.
    \item \textbf{Cursor:} We utilized version v1.5 (release date: August 21, 2025) of the standalone IDE.
\end{itemize}


\section{Prompt Templates}
\label{sec:appendix_prompts}
\begin{figure*}[t!]
    \centering
    \begin{subfigure}[b]{0.25\textwidth}
        \centering
        \fbox{\includegraphics[width=\textwidth]{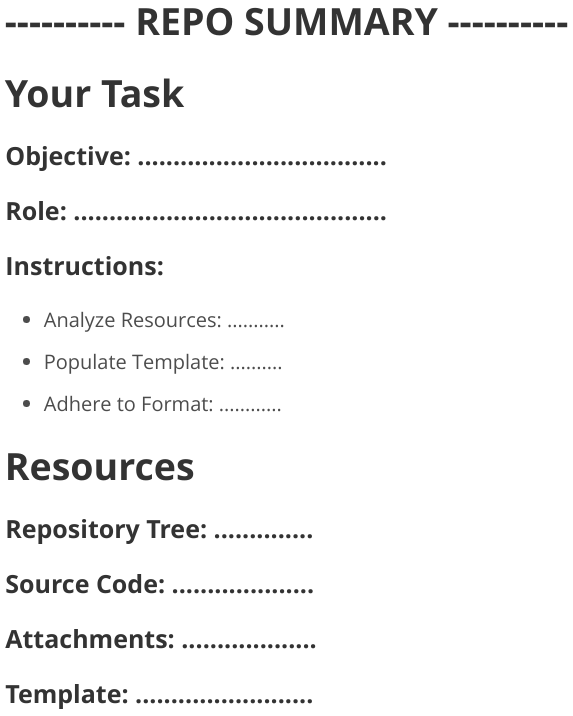}}
        \caption{Repository Summary Generation}
        \label{fig:repo-summary-prompt}
    \end{subfigure}
    \hfill
    \begin{subfigure}[b]{0.25\textwidth}
        \centering
        \fbox{\includegraphics[width=\textwidth]{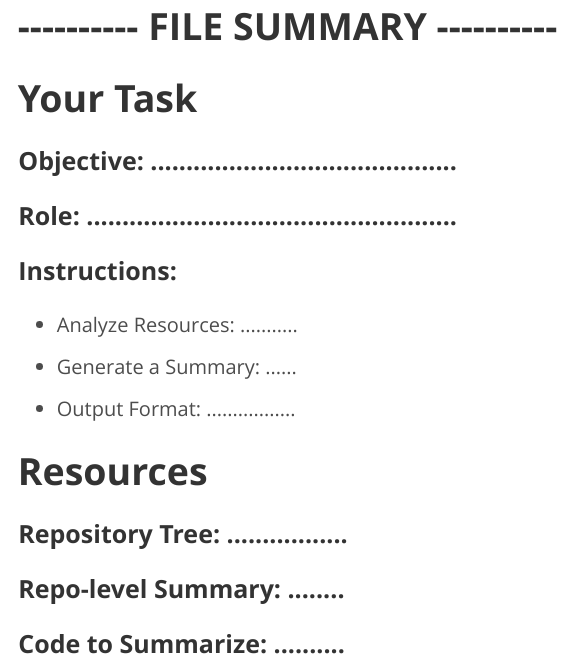}}
        \caption{File Summary Generation}
        \label{fig:file-summary-prompt}
    \end{subfigure}
    \hfill
    \begin{subfigure}[b]{0.25\textwidth}
        \centering
        \fbox{\includegraphics[width=\textwidth]{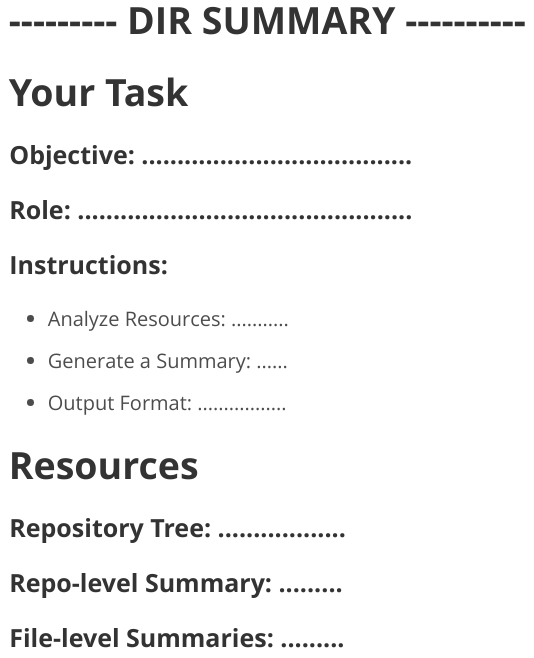}}
        \caption{Directory Summary Aggregation}
        \label{fig:dir-summary-prompt}
    \end{subfigure}
    
    \vspace{4mm} 
    
    \begin{subfigure}[b]{0.25\textwidth}
        \centering
        \fbox{\includegraphics[width=\textwidth]{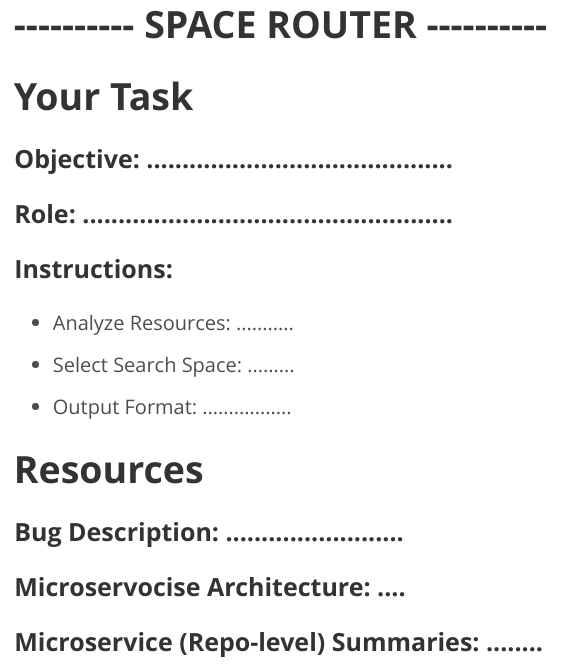}}
        \caption{Multi-Repository Routing}
        \label{fig:router-prompt}
    \end{subfigure}
    \hfill
    \begin{subfigure}[b]{0.25\textwidth}
        \centering
        \fbox{\includegraphics[width=\textwidth]{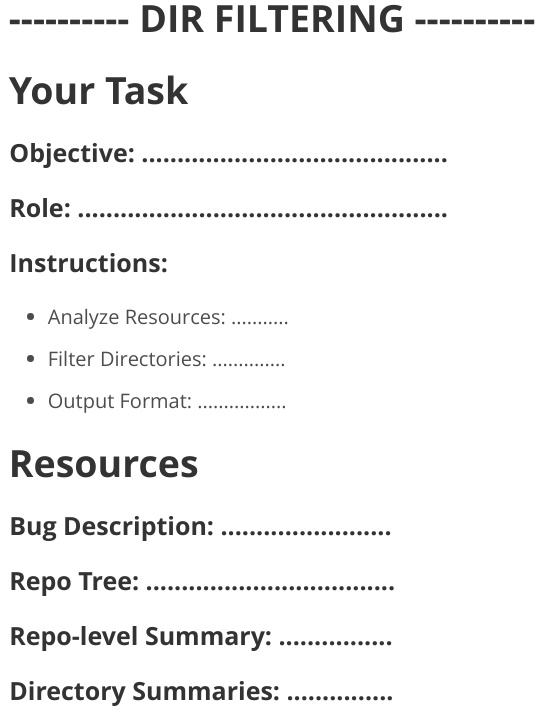}}
        \caption{Directory-Level Filtering}
        \label{fig:dir-filter-prompt}
    \end{subfigure}
    \hfill
    \begin{subfigure}[b]{0.25\textwidth}
        \centering
        \fbox{\includegraphics[width=\textwidth]{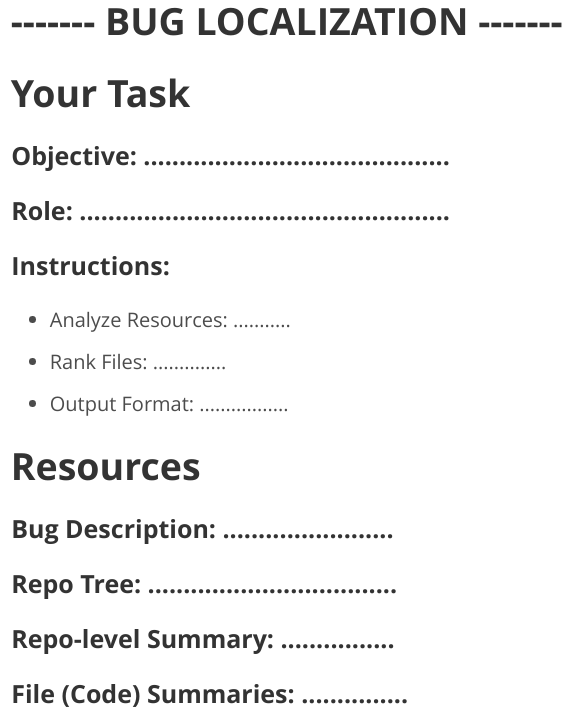}}
        \caption{File-Level Ranking}
        \label{fig:bug-loc-prompt}
    \end{subfigure}
    
    \caption{Core prompt templates used in our methodology. The top row (a-c) corresponds to the offline Knowledge Base generation phase, while the bottom row (d-f) corresponds to the online, inference-time bug localization phase.}
    \label{fig:all_prompts}
\end{figure*}

To provide further insight into our methodology, this section contains the core prompt templates used for both the offline knowledge base generation and the online bug localization phases. Figure~\ref{fig:all_prompts} illustrates the six key prompts that power our framework. These templates are simplified for clarity but represent the essential structure and inputs provided to the LLM at each stage of the process.

\section{Cost of Knowledge Base Construction}
\label{app:cost}

To summarize the entire project and generate the hierarchical NL knowledge base, we used \texttt{gpt-4.1-mini}. 
The total cost for summarizing all repositories was approximately \$80 (USD). 
While this is a manageable one-time cost in our industrial setting, it highlights the need for future research on \textbf{scheduling algorithms} to incrementally update and maintain summaries in a cost-efficient manner.

\end{document}